\documentclass[prl,twocolumn,showpacs,preprintnumbers,amsmath,amssymb]{revtex4}

\usepackage{graphicx}
\usepackage{amsmath}
\usepackage{epstopdf}
\usepackage{amsbsy}
\usepackage{dcolumn}
\usepackage{bm}

\newcommand{\BSCCO}{{Bi$_2$Sr$_2$CaCu$_2$O$_{8+x}$ }}

\begin{document}

\title{Breakdown of universal transport in correlated $d$-wave superconductors}

\author{Brian M. Andersen$^1$ and P. J. Hirschfeld$^2$}

\affiliation{$^1$Nano-Science Center, Niels Bohr Institute,
University of Copenhagen, Universitetsparken 5, DK-2100 Copenhagen,
Denmark\\
$^2$Department of Physics, University of Florida, Gainesville,
Florida 32611-8440, USA}

\date{\today}

\begin{abstract}

 The prediction and observation of low-temperature universal thermal conductivity
 in cuprates has served as a keystone of
 theoretical approaches to the superconducting state, but recent
 measurements on underdoped samples show strong violations of this
 apparently fundamental property of  $d$-wave nodal
 quasiparticles.  Here, we show that the breakdown of universality
 may be understood as the consequence of disorder-induced magnetic
 states in the presence of increasing antiferromagnetic
 correlations in the underdoped state, even as these same correlations
 protect the nodal low-energy density of states in agreement with recent scanning tunneling experiments.

\end{abstract}

 \maketitle

{\it Introduction.} Thermal conductivity measurements at low
temperatures $T$ in the superconducting state have played an
important role in strengthening the case for a $d$-wave BCS
description of quasiparticles in  optimally doped cuprate
superconductors. They are bulk probes of the superconducting
state, unlike angular resolved photoemission (ARPES) and scanning
tunneling microscopy (STM), and can currently be performed at
lower $T$ than microwave experiments.  One drawback is the need to
separate phonon and electron contributions, but in the cuprate
superconductors an asymptotic linear term, $\kappa_0/T\sim {\rm
const}$ which can be attributed solely to quasiparticles, dominates at the lowest $T$.  After theoretical
predictions of the universality of low $T$ quasiparticle transport
in nodal superconductors\cite{universal}, experimental
confirmation was obtained in optimally doped
materials\cite{universalexpt}.  According to theory, which relies
on the disorder-averaged self-consistent $T$-matrix approximation
(SCTMA), the low-$T$ thermal conductivity is
\begin{equation}
\kappa_{00}={\kappa_0\over T} \simeq {k_B^2\over 3\hbar} \left(
{{v_F\over v_\Delta} + {v_\Delta\over v_F} }\right), \label{eq1}
\end{equation}
where $k_B$ is Boltzmann's constant, and $v_\Delta$, $v_F$
denote the nodal gap slope and Fermi velocity,
respectively. This is a remarkable  example of a transport
coefficient which is unaffected--to leading order--by the addition
of disorder. Furthermore, this result is insensitive to vertex
corrections due to anisotropic impurity scattering and
Fermi liquid effects\cite{DurstLee}. Observation of
universal conductivity in optimally doped
samples\cite{MChiao:1999,SNakamae:2001} was a key step in
establishing the existence of nodal quasiparticles and the
validity of the SCTMA in this limit.  Soon thereafter, however,
measurements in underdoped samples exhibited values of the low-$T$
thermal conductivity considerably below predicted universal
limits\cite{exptlunderdoped}.  How and why the universal
prediction  (\ref{eq1}) breaks down has never been explained.

Since $v_F$ is generally considered to be well-known from ARPES\cite{ARPES},
Eq.(\ref{eq1}) has also been used to extract the gap slope for a
number of cuprates at lower doping as well\cite{exptlunderdoped},
leading to the conclusion that $v_\Delta$ increases with
underdoping\cite{exptlunderdoped}. This conclusion is in apparent
contradiction to  recent Raman\cite{Raman} and ARPES
experiments\cite{ARPES}, so it is even more important to examine physical
effects outside the framework of the SCTMA which could lead to a
suppression of $\kappa_0/T$ and thereby to a possible erroneous
conclusion about the doping dependence of $v_\Delta$.

There are several effects known to lead to a suppression of
$\kappa_0/T$. Localization effects were discussed in this context
in Ref. \cite{Atkinson}, and effects of bulk subdominant competing
orders have been shown to suppress $\kappa_0/T$ but do not
immediately eliminate it, despite the removal of the $d$-wave
nodes\cite{Gusynin}. Here, we investigate the effects on
$\kappa(T)$ by local impurity-induced moments relevant e.g. to the
underdoped regime of La$_{2-x}$Sr$_x$CuO$_4$ (LSCO)\cite{bella,julien,BMAndersen:2006}.
Such moments are formed in the presence of background Hubbard
interactions in the host material\cite{HAlloul:2007}.

Recently, the effect of strong correlations of this type on the
density of states (DOS) of a disordered $d$-wave superconductor was
investigated by Garg {\it et al.} \cite{garg}, in an approach where the
Gutzwiller approximation was used to approximately project out the
doubly occupied states of a disordered $t-J$ model.  They found
the rather remarkable result that the low-energy states were ``protected" by interactions, i.e. in
the presence of the projection the low-energy residual DOS which normally arises from disorder\cite{balatsky}
was strongly suppressed. Experimentally it is indeed observed that the low-energy DOS is surprisingly 
homogeneous\cite{pan,mcelroy,vershinin}, reflecting a robustness of the nodal quasi-particles to disorder.  Garg {\it et al.}
speculated that the projection mitigated effects of disorder
{\it generally} at low energies. Here we show that  in the density
channel this hypothesis is correct: interactions screen and
diminish the disorder potential.  In the
magnetic channel, however, scattering is enhanced, and properties
which are sensitive to spin scattering like the universal thermal
conductivity $\kappa(T)$ may be strongly renormalized even in the absence of Anderson localization.

{\it Model.} The model we use to study disordered $d$-wave
superconductors with magnetic correlations is
\begin{eqnarray}\label{Hamiltonian}
\hat H = &-&\sum_{\langle ij \rangle \sigma}t_{ij}\hat
c_{i\sigma}^\dagger\hat c_{j\sigma} + \sum_{i\sigma} \left(
Un_{i,{-\sigma}} + V^{imp}_i - \mu\right)\hat
c_{i\sigma}^\dagger\hat c_{i\sigma}
\nonumber\\
&+& \sum_{\langle ij\rangle}\left( \Delta_{ij}\hat
c_{i\uparrow}^\dagger \hat c_{j\downarrow}^\dagger +
\mbox{H.c.}\right).
\end{eqnarray}
Here, $\hat{c}_{i\sigma}^\dagger$ creates an electron on site $i$
with spin $\sigma$, and $t_{ij}=\{t,t'\}$ denote the two nearest
neighbor hopping integrals, $V^{imp}_i=\sum_{j=1}^P V^{imp}
\delta_{ij}$ is a nonmagnetic impurity potential resulting from a
set of point-like scatterers of strength $V^{imp}$, $\mu$ is
the chemical potential adjusted to fix the doping $x$, and
$\Delta_{ij}$ is the
order parameter on the bond between sites $i$ and $j$. The
amplitude of $\Delta_{ij}$ is set by the coupling constant
$g/t=1.3$. The Hubbard repulsion is treated by
an unrestricted Hartree approximation, and $U$ is taken to be homogeneous.
Below we fix $t'/t=-0.3$, $x=0.1$, and solve
Eq.(\ref{Hamiltonian}) self-consistently by diagonalizing the
associated Bogoliubov-de Gennes (BdG) equations on systems of
$N=40\times 40$ sites\cite{JWHarter:2006,andersen07}.

The model given by Eq.(\ref{Hamiltonian}) has been used
extensively in the literature to study bulk competing phases,
field-induced magnetization, as well as novel bound states at
interfaces between antiferromagnets and
superconductors\cite{allHamiltonian}. It has also been used to
study field-induced moment formation around nonmagnetic impurities
in correlated $d$-wave
superconductors\cite{JWHarter:2006,YOhashi:2002,YChen:2004}. In
the case of many impurities, Eq.(\ref{Hamiltonian}) was recently
used to model static disorder-induced antiferromagnetism as seen,
e.g., by  neutron scattering measurements\cite{bella,andersen07}.

{\it Results.} In the clean case ($V^{imp}=0$), magnetic order induced by $U$
will compete with the superconducting order and lead
to a bulk magnetic state above a critical value $U_{c2}$. For
$U_{c1}<U<U_{c2}$, a single point-like impurity can induce a localized
$S=1/2$ state with staggered
magnetization\cite{YOhashi:2002,YChen:2004,JWHarter:2006,ZWang:2002}. Here,
$U_{c1}$ is the critical $U$ necessary for local moment formation,
and depends on band structure and impurity strength. When
$U<U_{c1}$ there is no induced magnetization but the magnetic
correlations still suppress the charge modulations and shorten the
superconducting healing length $\xi$ near the impurity.

The picture becomes considerably more complex for finite impurity
concentrations. However, there is still a lower value $U_{c1}^*$
below which no magnetism is induced, and an upper $U_{c2}^*\simeq
U_{c2}$ above which the ground state becomes that of a disordered
magnet. As explained in Ref. \onlinecite{andersen07}, it  easier
for a dirty system to generate antiferromagnetic islands
($U_{c1}^* < U_{c1}$), because of clustering of impurities and
resulting charge redistributions.

{\it Density of states.}   We first focus on the DOS of the disordered $d$-wave
superconductor in the presence of correlations and compare with
the results of Ref. \cite{garg}.  In particular, we would like to
ascertain whether the onset of the local magnetic state has a
qualitative effect on the total DOS.
\begin{figure}[b]
\begin{center}
\leavevmode \begin{minipage}{.49\columnwidth}
\includegraphics[clip=true,height=.8\columnwidth,width=1.1\columnwidth]{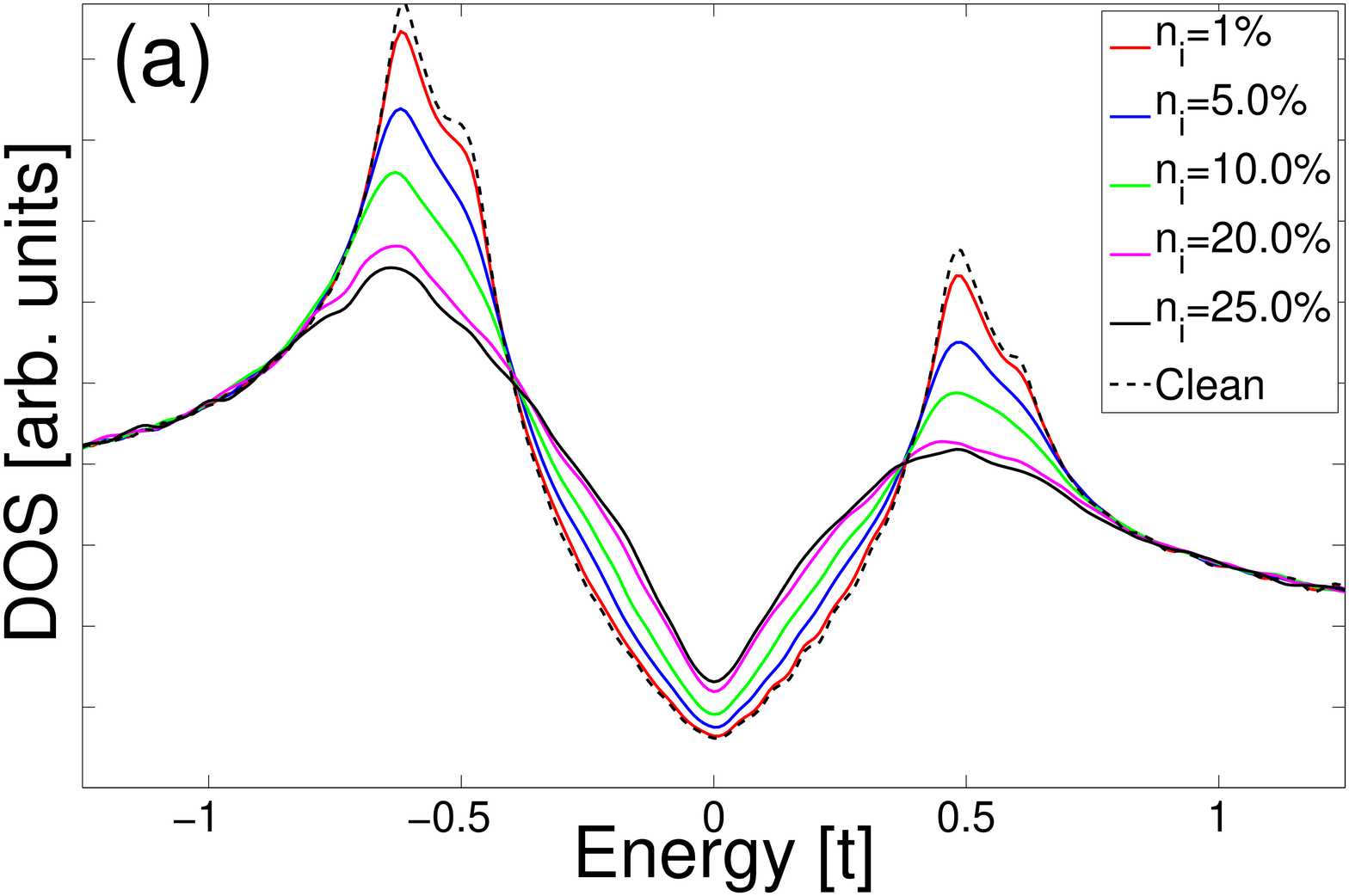}
\end{minipage}
\begin{minipage}{.49\columnwidth}
\includegraphics[clip=true,height=.8\columnwidth,width=1.1\columnwidth]{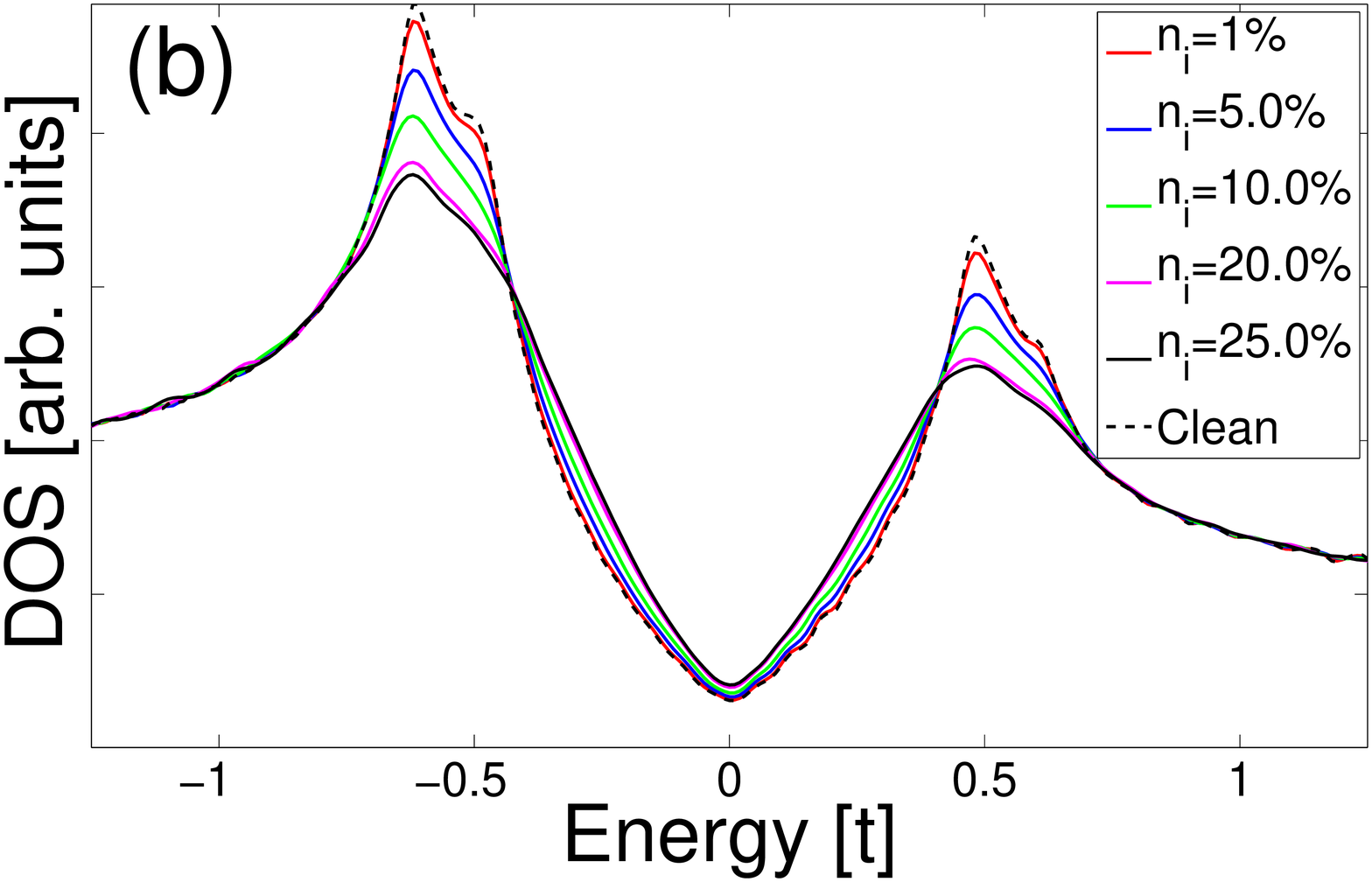}
\end{minipage}
\begin{minipage}{.49\columnwidth}
\includegraphics[clip=true,height=.8\columnwidth,width=1.1\columnwidth]{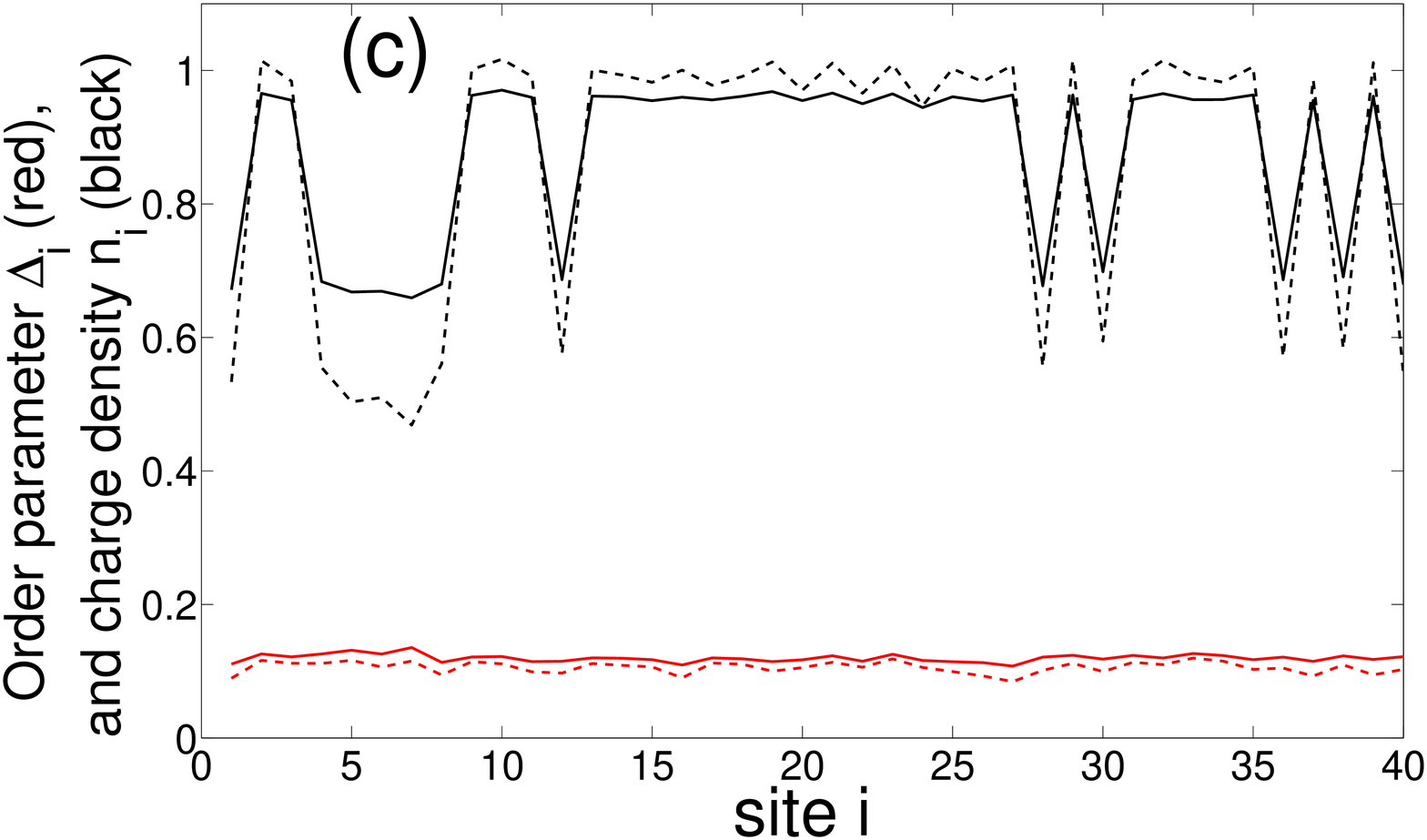}
\end{minipage}
\begin{minipage}{.49\columnwidth}
\includegraphics[clip=true,height=.8\columnwidth,width=1.1\columnwidth]{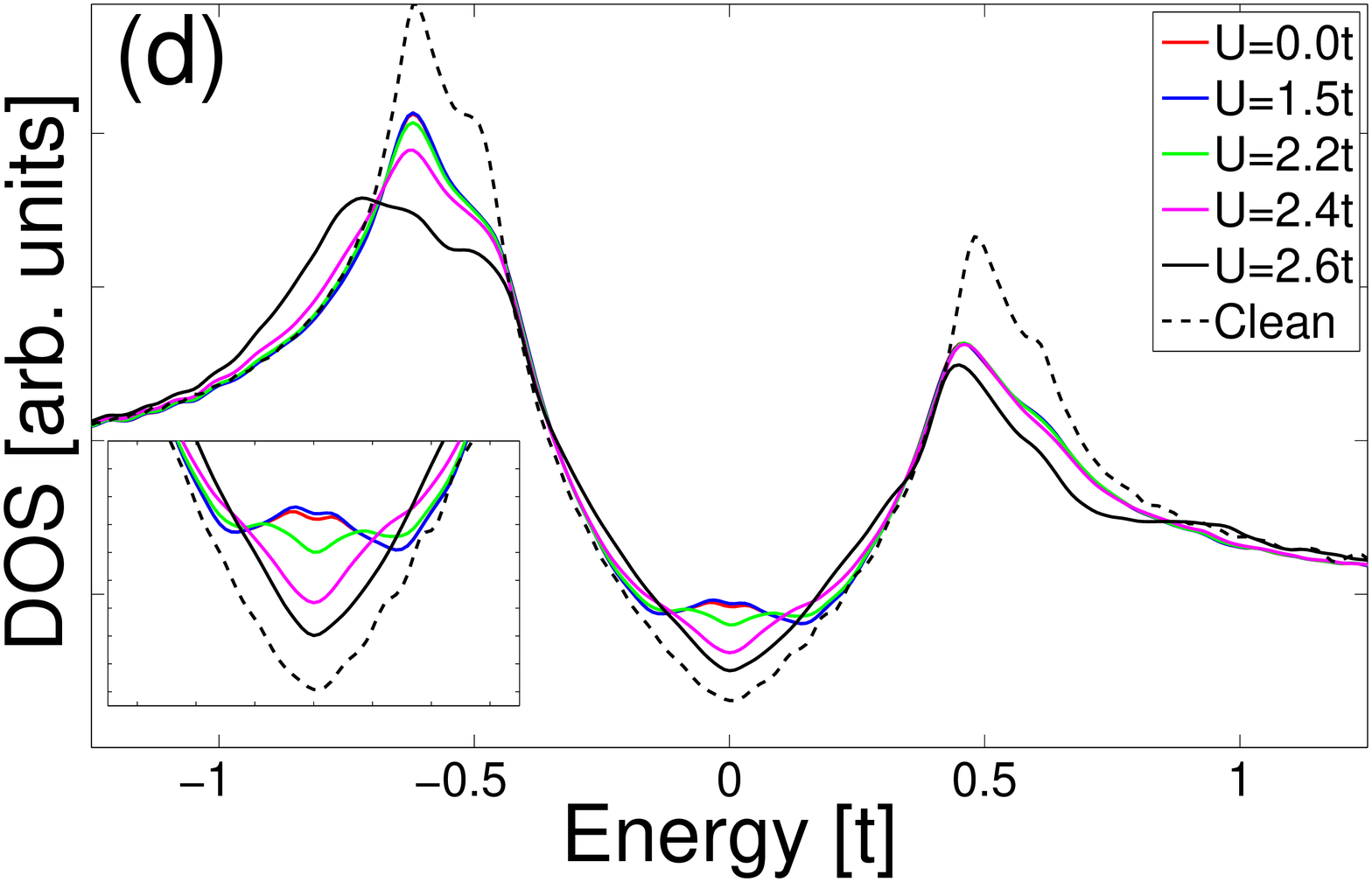}
\end{minipage}
\caption{(Color online) (a) Spatially averaged DOS in the uncorrelated $d$-wave
superconductor ($U=0$) for different concentrations of disorder
$n_i$. One clearly sees the pile-up of low-energy impurity states.
(b) Same as (a) but for the correlated $d$-wave
superconductor with $U=2.5t$. For a large range of impurity
concentrations (0-25\%) any sizable and qualitative change in the 'V'-shaped
low-energy DOS is completely absent. (c) A line cut through the
center of the system showing the total charge density (top black
lines) and the $d$-wave order parameter (lower red lines).
The dashed (solid) lines display the results without (with)
correlations.  (d) DOS for 2.5\% strong scatterers ($V^{imp}/t=10$) as a function of $U$. The inset shows a zoom of the low-energy region $\omega \in [-0.35t,0.35t]$.} \label{fig:DOS}
\end{center}
\end{figure}
In Fig. \ref{fig:DOS}a, we show the spatially averaged total DOS,
$N(\omega)=\sum_{{\mathbf{r}_i}} N({\mathbf{r}_i},\omega)/N$,
where
\begin{equation}
N({\mathbf{r}_i},\omega)=\sum_{n\sigma} |u_{n\sigma}(i)|^2 \delta(\omega-E_{n\sigma}) +
|v_{n\sigma}(i)|^2 \delta(\omega+E_{n\sigma})\label{DOS},
\end{equation}
for a series of disorder concentrations consisting of weak
scatterers $V^{imp}/t=1.0$ for the noninteracting case $U=0$. We
see the usual low-energy pile-up of impurity states inside the
$d$-wave gap\cite{atkinson2}. In order to obtain smooth DOS curves
we have used a small artificial smearing factor $\eta/t=0.025$.
Figure \ref{fig:DOS}b displays the results of the same study, but
for the correlated $d$-wave superconductor with $U/t=2.5$. The
remarkable result is that even for 25\% disorder, the Hubbard
correlations "protect" the $d$-wave V-shaped DOS\cite{garg}. The
origin of this universal low-energy behavior is the suppressed
charge modulations near the impurities as shown in Fig. \ref{fig:DOS}c. Therefore, an
impurity potential apparently disturbs a correlated superconductor
much less than a conventional BCS state\cite{garg}. Interestingly, impurities which only modulate the
pair interaction will also perturb the electronic structure
primarily near the antinodes, while protecting the low-energy DOS
universality. This kind of disorder may exist in \BSCCO, where dopants seem to enhance the pairing interaction
locally\cite{nunner1,andersentherm}.

The results shown in Fig. \ref{fig:DOS}a-c are for weak scatterers
without induced magnetization, i.e. in the regime $U < U_{c1}^*$.
It is also interesting to study the DOS in the cluster spin-glass
phase, $U_{c1}^* < U < U_{c2}^*$, highly relevant for example for
underdoped LSCO\cite{bella,julien,andersen07}.  The tendency to
create impurity-induced magnetic states increases with the
impurity potential\cite{HAlloul:2007}, so in Fig. \ref{fig:DOS}d
we have compared the DOS with near-unitarity limit scatterers with
$V^{imp}/t=10$ and a fixed impurity concentration of $n_i=2.5\%$.
From Fig. \ref{fig:DOS}d it is evident that in the absence of
correlations, the conventional plateau in the DOS predicted by the
SCTMA is seen. As $U$ is increased, this plateau disappears
gradually, and there is no qualitative change in the DOS as
$U^*_{c1}/t\simeq 1.5$ is crossed. Note at this relatively weak
value of $U$, the projection of doubly occupied sites has not been
fully implemented, but the tendency is clearly the same.
As $U$ increases beyond $U_{c2}/t\simeq
U^*_{c2}/t\simeq2.8$ we enter a regime where bulk magnetic order coexists with superconductivity, and the
clean DOS acquires an additional gap of the Mott-Hubbard type. However, as shown recently, even in this regime, disorder can result in low-energy DOS indistinguishable from a $d$-wave superconductor\cite{atkinson2007}.

{\it Thermal conductivity.} For $d$-wave superconductors, a real-space BdG calculation of
$\kappa(T)$ was used to study both  localization
phenomena\cite{Atkinson, BMAndersen:2006} and the vortex
state\cite{takigawa}. Here we follow the same approach but focus
on the correlation effects and the role of disorder-induced
magnetization using a realistic band structure for cuprate
superconductors. For an inhomogeneous system we define the total
thermal conductivity $\kappa(T)$ as
\begin{equation}
\kappa(T)\equiv\kappa_{xx}(T)=\frac{1}{N}\sum_{\mathbf{r}_i}\kappa_{xx}({\mathbf{r}_i},T),
\end{equation}
where $\kappa_{xx}({\mathbf{r}_i},T)=h_x({\mathbf{r}_i})/(-\nabla_x
T)$ is the ratio of the thermal current along the $x$ axis at position
${\mathbf{r}_i}$ and the uniform temperature gradient applied along
the $x$ direction. Within linear response we have
\begin{equation} \kappa(T)=\frac{1}{T} \mbox{Im} \left[
\frac{d}{d\Omega} \frac{1}{N}\sum_{{\mathbf{r}_i},{\mathbf{r}_j}}
Q_{xx}({\mathbf{r}_i},{\mathbf{r}_j},i\Omega_n\rightarrow
\Omega+i0^+) \right]_{\Omega\rightarrow 0}\label{linreskap},
\end{equation}
where $Q_{xx}$ is the heat current-heat current correlation
function. The quantity inside the parenthesis in Eq.\eqref{linreskap} reads in compact form
\begin{widetext}
\begin{equation}\label{kernel}
\frac{d}{d\Omega} \frac{1}{N}\sum_{{\mathbf{r}_i},{\mathbf{r}_j}}
Q_{xx}({\mathbf{r}_i},{\mathbf{r}_j},i\Omega_n\rightarrow
\Omega+i0^+) |_{\Omega\rightarrow 0}=
\frac{1}{4N}\sum_{n,m}F(E_n,E_m)
 |\sum_{{\mathbf{r}_i},{\mathbf{r}_l}} \psi_n^*(i) \hat{v}_g \psi_m(l)|^2,
\end{equation}
where the vector $\psi_n(i)=[u_n(i),v_n(i)]$, and $\hat{v}_g$ denotes the
discretized group velocity operator given by
$\hat{v}_g=\hat{v}^x_{kin}\tau_3+\hat{v}^x_\Delta\tau_1$\cite{Atkinson}.
Here the kinetic velocity
$\langle\hat{v}^x_{kin}\rangle_{il}=-it_{il}(x_i-x_l)$ with
$t_{il}=t(t'/\sqrt{2})$ for $i,l$ being nearest(next-nearest)
neighbors (in the flow direction), and similarly the gap velocity is given by
$\langle\hat{v}^x_{\Delta}\rangle_{il}=i\Delta_{il}(x_i-x_l)$. In
Eq.\eqref{kernel} we have introduced the spatially independent
thermal function $F(E_n,E_m)$ given by
\begin{equation}
F(E_n,E_m)=\int\frac{d\omega}{2\pi}\int\frac{d\omega'}{2\pi}
\delta_\eta(\omega-E_n)\delta_\eta(\omega'-E_m) \left[ P
\frac{\omega^2f(\omega)-\omega'^2f(\omega')}{(\omega-\omega')^2} +
i\pi\omega^2 f'(\omega)\delta(\omega-\omega')\right]\label{ffct},
\end{equation}
\end{widetext}
where $\delta_\eta(\omega)=2\eta/(\omega^2+\eta^2)$\cite{takigawa}. The spin index
(and the sum over spin) is implicit in Eqs.\eqref{kernel}-\eqref{ffct}.

Numerical solution of Eqs.\eqref{kernel}-\eqref{ffct} is
computationally demanding, and we are restricted to systems
of order $40\times 40$ sites. To study the low $T$ regime, one
needs to have a sufficient number of states within the $d$-wave
gap. Therefore, for the discussion below we have increased the
coupling constant to $g/t=2.8$, giving $\Delta_{ij}/t=0.4$ per
link in the homogeneous case. This value of $\Delta/t$ gives a
universal value $\kappa_{00}=1/3\left( {{v_F/v_\Delta} +
{v_\Delta/v_F} }\right)=0.88$ ($k_B=\hbar=1$) for the band used here. We
stress that the increase of the superconducting gap, which is a
factor 6-8 larger than cuprate materials, is merely a means to
calculate $\kappa(T)/T$ at low $T$ and is not important for the
discussion below. However, the
corresponding critical $U$'s will be larger as well (as discussed
below).

\begin{figure}[t]
\begin{center}
\leavevmode \begin{minipage}{.49\columnwidth}
\includegraphics[clip=true,height=0.8\columnwidth,width=1.1\columnwidth]{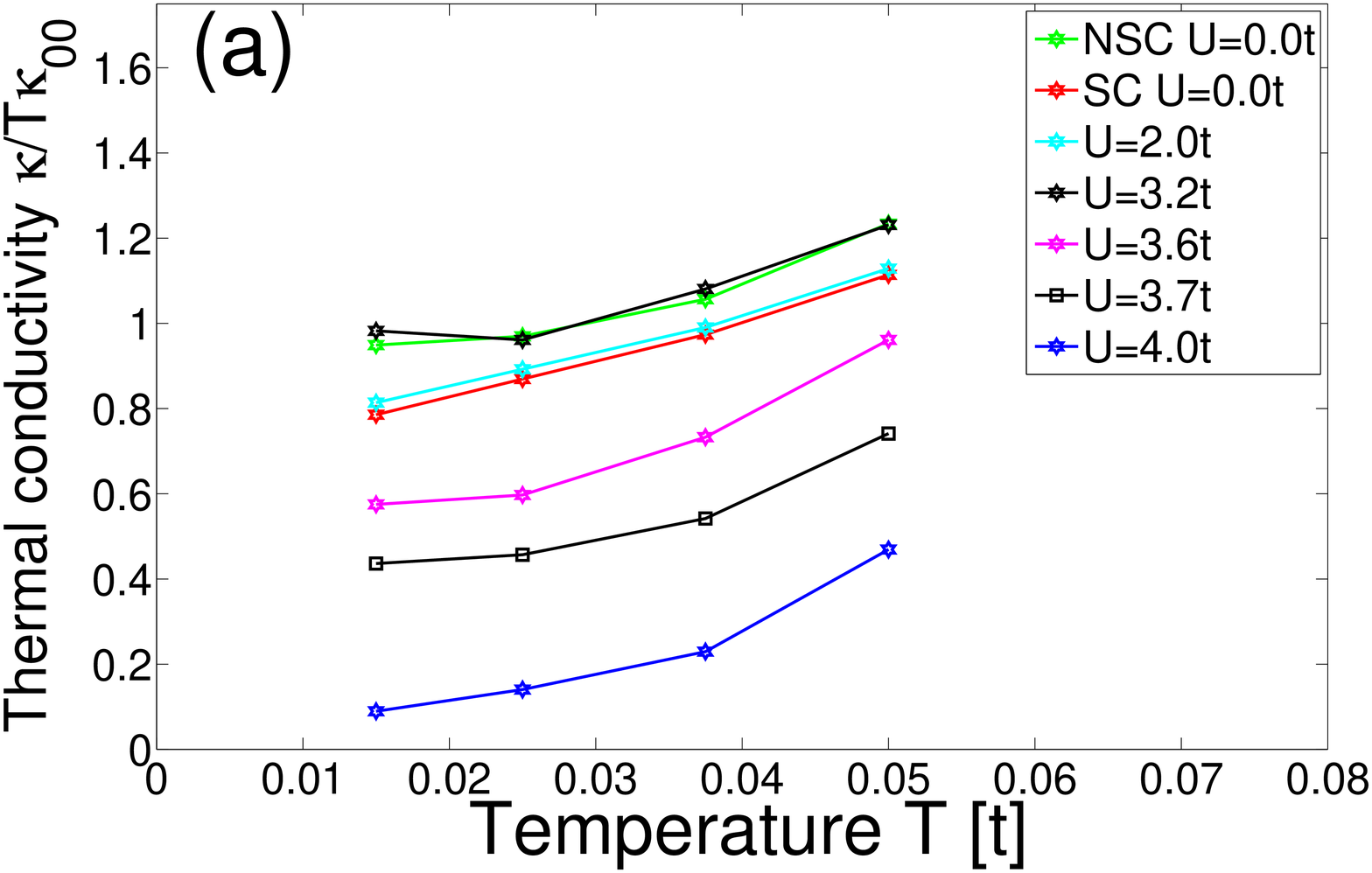}
\end{minipage}
\begin{minipage}{.49\columnwidth}
\includegraphics[clip=true,height=0.8\columnwidth,width=1.1\columnwidth]{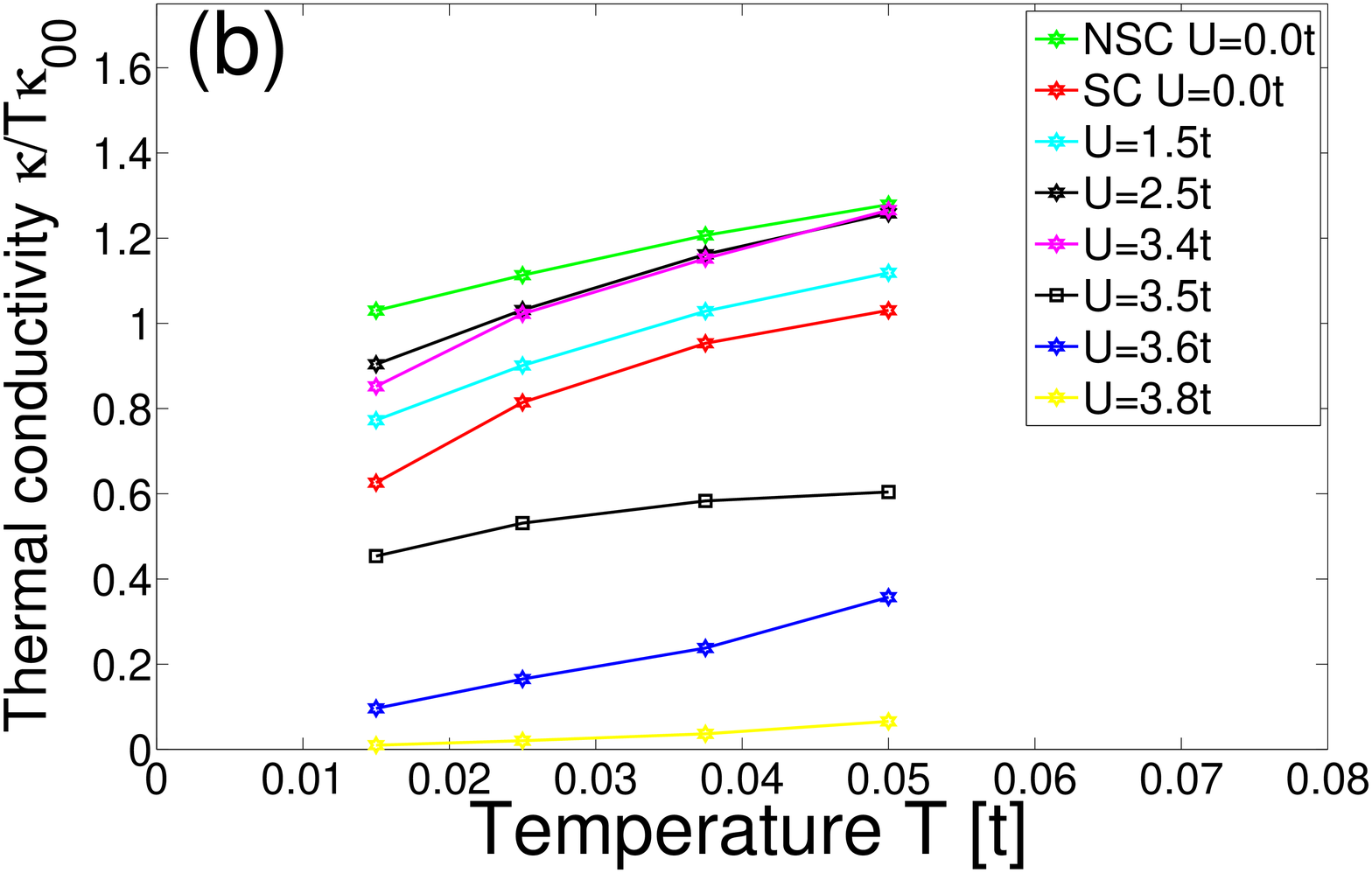}
\end{minipage}
\begin{minipage}{.49\columnwidth}
\includegraphics[clip=true,height=0.8\columnwidth,width=1.1\columnwidth]{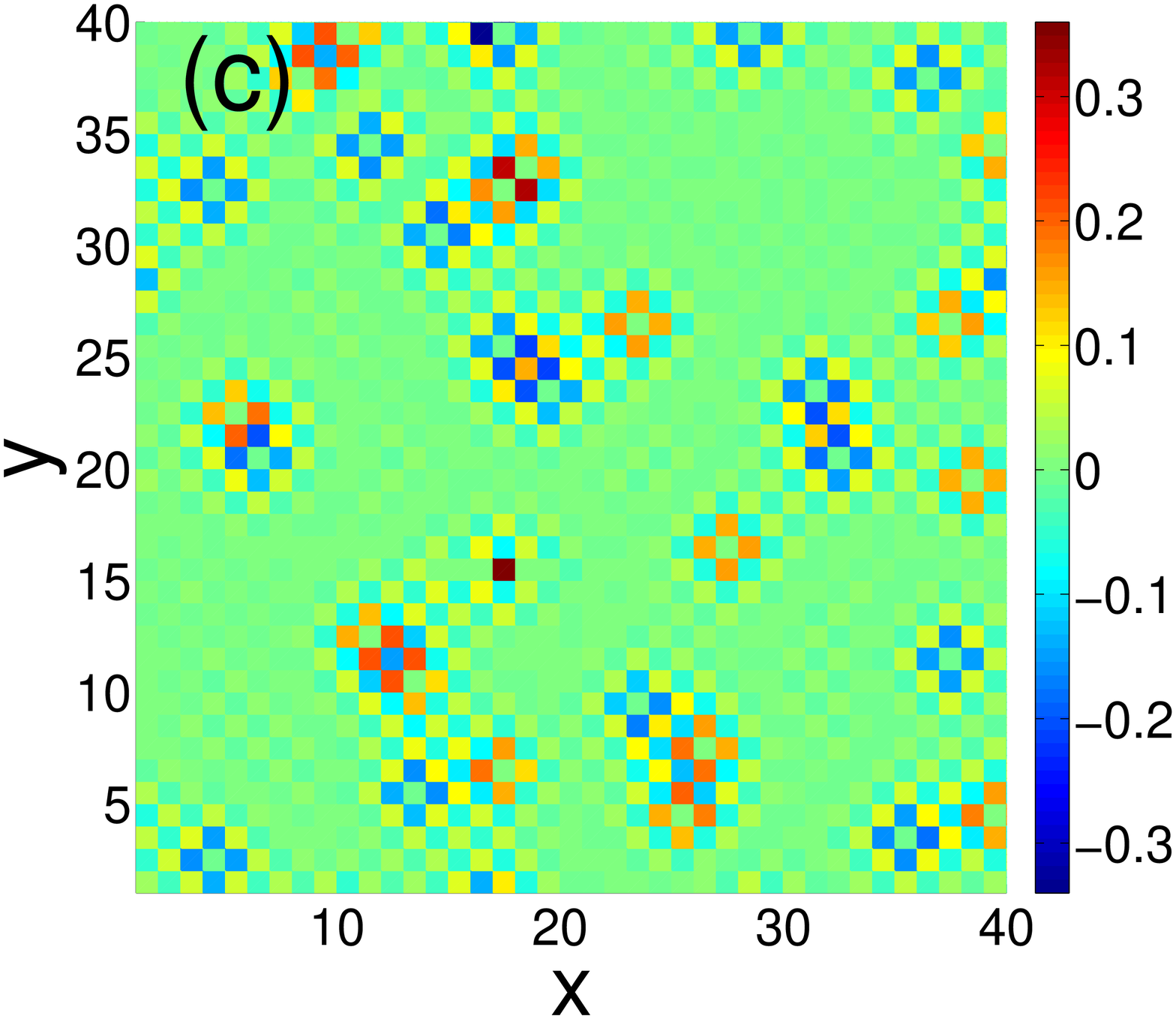}
\end{minipage}
\begin{minipage}{.49\columnwidth}
\includegraphics[clip=true,height=0.8\columnwidth,width=1.1\columnwidth]{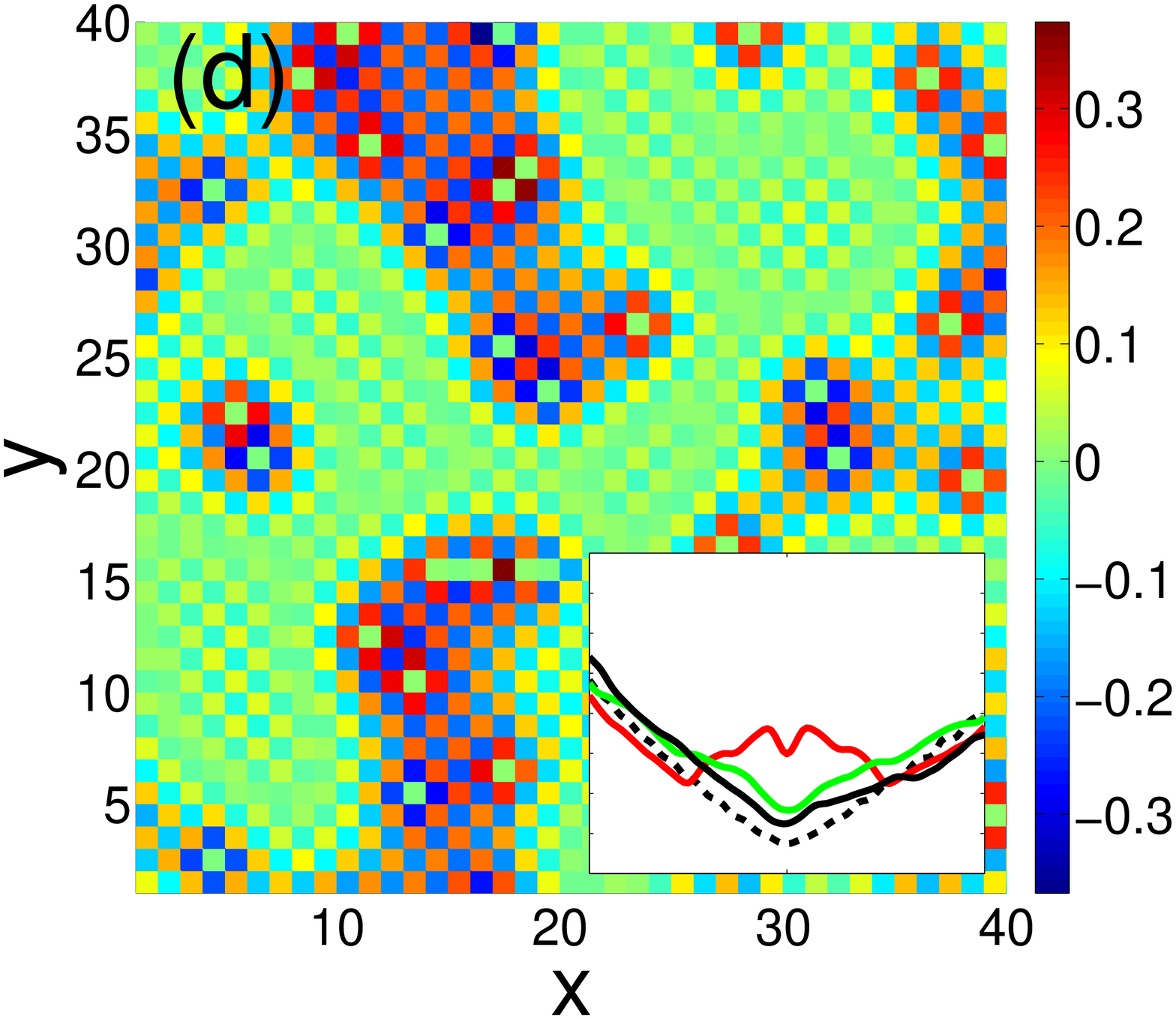}
\end{minipage}
\caption{(Color online) (a-b) $\kappa(T)/T$ versus $T$ as a function of increased $U$. (a) shows the case of 2.5\% strong scatterers with $V^{imp}/t=10.0$, and (b) has 15.0\% weaker impurities of $V^{imp}/t=3.0$. (c-d) show the disorder-induced magnetization in real-space for the case in (a) with $U/t=3.2$ (c) and $U/t=3.6$ (d). The inset in (d) shows the DOS ($\omega \in [-0.3t,0.3t]$) for the clean system (dashed), and the disordered case from (a) with $U=0$ (red), $U/t=3.2$ (green), and $U/t=3.6$ (black). } \label{kappafig}
\end{center}
\end{figure}

In Fig. \ref{kappafig}a (Fig. \ref{kappafig}b), we show the
low-$T$ thermal conductivity for the case of $n_i=2.5\%$ ($15\%$)
scatterers with $V^{imp}/t=10$ ($V^{imp}/t=3$) averaged over 20
different random impurity configurations (which is enough for
configurational convergence). For $U=0$ the result in
\ref{kappafig}a agrees well with those obtained previously for the
dilute impurity limit: in the non-self-consistent (NSC)
calculation (with homogeneous $\Delta$=0.4), $\kappa(T)/T = \kappa_{00} +
\alpha T^2$\cite{PJHirschfeld:1996}, whereas in the
self-consistent (SC) case the spatial inhomogeneity of
$\Delta_{ij}$ causes a reduction of $\alpha$ and leads to a more
linear $T$ dependence\cite{Atkinson} of $\kappa/T$. The case with
$V^{imp}/t=3.0$ in \ref{kappafig}b is an example of one of the
weakest disorder potentials we are able to study since further
reduced $V^{imp}$ leads to mean-free-path larger that our
system. As seen from Figs. \ref{kappafig}a-b, an
increase of the correlations $U$, initially increases
$\kappa(T)/T$ towards the NSC result. The origin of this
enhancement is the $U$-suppression of the charge- and gap-modulations: the self-consistent mean fields approach the NSC
result. As $U$ increases further and magnetic scatterering centers
are induced near the defects (Fig. \ref{kappafig}c-d),
$\kappa(T)/T$ is continuously suppressed, and $\kappa_0/T$
eventually vanishes in the bulk magnetic state.
In BCS $d$-wave superconductors the universal ratio of $\kappa_0/T$ arises from a 
cancellation of the increased disorder-induced scattering rate and a concomitant increase in the density 
of low-energy quasi-particles. In the present case, the universality is broken because the disorder-induced 
moments increase the scattering rate whereas the resulting low-energy DOS remains unchanged. 

In the two cases shown in Fig. \ref{kappafig}a-b,
$U^*_{c1}/t\simeq 1.9$ for $V^{imp}/t=10.0$, and $U^*_{c1}/t\simeq
3.1$ for $V^{imp}/t=3.0$, whereas $U_{c2}/t=3.9$ in both cases.
Density of states corresponding to the magnetization plots in Fig.
\ref{kappafig}c-d are shown in the inset of Fig. \ref{kappafig}d.
Thus, there exists a large spin-glass regime where the DOS remains
universal in the sense that it maintains the characteristic
'V'-shape gap for a $d$-wave superconductor, but the thermal
conductivity is simultaneously suppressed due to the creation of
additional effective magnetic scatterers. Therefore,
impurity-driven local moment formation present in the underdoped
regime may explain the doping dependence of $\kappa_0/T$ measured
in LSCO\cite{takeya}. Since the low-$T$ thermal conductivity is
strongly suppressed, a naive use of the "universal" clean $d$-wave
result for $\kappa_{00}$ in Eq. (\ref{eq1}) would lead to an
erroneous estimate of the superconducting gap slope at the node.
Thus the conclusion that $v_\Delta$ is reduced with decreasing
doping as found by recent  Raman\cite{Raman} and ARPES\cite{ARPES}
measurements is not necessarily inconsistent with thermal
conductivity measurements\cite{exptlunderdoped,takeya} within the
picture presented here. Note that since the mean field theory
employed here overestimates the tendency towards static magnetic
order, our results suggest that the thermal conductivity in the
presence of low-energy dynamical spin fluctuations not quite
pinned by impurities, as apparently true in YBCO, will also be
suppressed relative to the universal limit.

In summary, we studied the effects of disorder in correlated
$d$-wave superconductors, and calculated the DOS and thermal
conductivity in different regimes of the Hubbard repulsion $U$.
Although the low-energy DOS is protected due to
screening of the disorder by interactions,  disorder
induced magnetism in the presence of correlations can lead to
enhanced scattering which may strongly modify transport
properties, in agreement with experiments on thermal conductivity
in underdoped cuprates.

We acknowledge discussions with K. Behnia, G. Boyd, P. Hedeg\aa rd,
A. P. Kampf, N. Trivedi. P. J. H. acknowledges support by DOE
Grant DE-FG02-05ER46236.


\begin{thebibliography}{00}

\bibitem{universal} P. A. Lee, Phys. Rev. Lett. {\bf 71},
1887 (1993); M. J. Graf {\sl et al.}, Phys. Rev. B {\bf 53}, 1514
(1996).
%
\bibitem{universalexpt} L. Taillefer {\sl et al.}, Phys. Rev. Lett. {\bf 79},
483 (1997).
%
\bibitem{DurstLee} A. C. Durst and P. A. Lee, Phys. Rev. B {\bf 62},
1270 (2000).
%
\bibitem{MChiao:1999} M. Chiao  {\sl et al.}, Phys. Rev. Lett. {\bf 82}, 2943 (1999).
%
\bibitem{SNakamae:2001} S. Nakamae  {\sl et al.}, Rev. B {\bf 63}, 184509 (2001).
%
\bibitem{exptlunderdoped} M. Sutherland {\sl et al.}, Phys. Rev. Lett. {\bf 94},
147004 (2005); X. F. Sun {\sl et al.}, Phys. Rev. Lett. {\bf 96},
017008 (2006).
%
\bibitem{Raman} M. Le Tacon {\sl et al.}, Nature Phys. {\bf 2}, 537 (2006).
%
\bibitem{ARPES} J. Mesot {\sl et al.}, Phys. Rev. Lett. {\bf 83},
840 (1999); K. Tanaka {\sl et al.}, Science {\bf 314}, 1910 (2006);
T. Kondo {\sl et al.}, Phys. Rev. Lett {\bf 98}, 267004 (2007).
%
\bibitem{Atkinson} W. A. Atkinson and P. J. Hirschfeld, Phys. Rev. Lett. {\bf 88},
187003 (2002).
%
\bibitem{Gusynin} V. P. Gusynin and V. A. Miransky, Eur. Phys.
J. B {\bf 37}, 363 (2004).
%
\bibitem{bella} B. Lake {\sl et al.}, Nature (London) {\bf 415}, 299 (2002).
%
\bibitem{julien} M.-H. Julien, Physica B {\bf 329-333}, 693  (2003).
%
\bibitem{BMAndersen:2006} B. M. Andersen and P. J. Hirschfeld, Physica (Amsterdam) {\bf 460C}, 744
(2007).
%
\bibitem{HAlloul:2007} H. Alloul, J. Bobroff, M. Gabay and  P. J. Hirschfeld,
arXiv:0711.0877v1.
%
\bibitem{garg} A. Garg, M. Randeria, and N. Trivedi,
arXiv:cond-mat/0609666.
%
\bibitem{pan} S. H. Pan, {\sl et al.}, Nature (London) {\bf 413}, 282 (2001).
%
\bibitem{mcelroy} K. McElroy, {\sl et al.}, Nature (London) {\bf 422}, 592 (2003); Phys. Rev.
Lett. {\bf 94}, 197005 (2005); Science {\bf 309}, 1048 (2005).
%
\bibitem{vershinin} M. Vershinin  {\sl et al.}, Science {\bf 303}, 1048 (2004).
%
\bibitem{balatsky} A.V. Balatsky, I. Vekhter and J.-X. Zhu, Rev. Mod. Phys. {\bf 78}, 373 (2006).
%
\bibitem{JWHarter:2006} J. W. Harter {\sl et al.}, Phys. Rev. B {\bf 75}, 054520 (2007).
%
\bibitem{andersen07} B. M. Andersen {\sl et al.}, Phys. Rev. Lett. {\bf 99}, 147002 (2007).
%
\bibitem{allHamiltonian} I. Martin {\sl et al.}, Int. J. Mod. Phys. {\bf 14}, 3567-3577
(2000); M. Ichioka, M. Takigawa, and K. Machida, J. Phys. Soc. Jpn.
{\bf 70}, 33 (2001); J.-X. Zhu, I. Martin, and A. R. Bishop, Phys. Rev. Lett. {\bf
89}, 067003 (2002); B. M. Andersen and P. Hedeg\aa rd, {\it ibid.} {\bf 95}, 037002 (2005); H.-Y. Chen and C. S. Ting, Phys. Rev
B {\bf 71}, 220510(R) (2005); B. M. Andersen {\sl et al.}, {\it ibid.} {\bf 72}, 184510 (2005).
%
\bibitem{YOhashi:2002} Y. Ohashi, Phys. Rev. B {\bf 66}, 054522 (2002).
%
\bibitem{YChen:2004} Y. Chen and C. S. Ting, Phys. Rev. Lett. {\bf 92}, 077203 (2004).
%
\bibitem{ZWang:2002} H. Tsuchiura {\sl et al.}, Phys. Rev B {\bf 64}, 140501(R) (2001); Z. Wang and P. A. Lee, Phys. Rev. Lett. {\bf 89}, 217002 (2002).
%
\bibitem{atkinson2} W. A. Atkinson, P. J. Hirschfeld, and A. H. MacDonald,
Phys. Rev. Lett. \textbf{85}, 3922 (2000).
%
\bibitem{nunner1} T. S. Nunner {\sl et al.}, Phys. Rev. Lett. {\bf 95}, 177003 (2005);  Phys. Rev. B {\bf 73}, 104511 (2006).
%
\bibitem{andersentherm} B. M. Andersen {\sl et al.}, Phys. Rev. B {\bf 74}, 060501(R) (2006).
%
\bibitem{atkinson2007} W. A. Atkinson, Phys. Rev. B {\bf 75}, 024510 (2007).
%
\bibitem{takigawa} M. Takigawa, M. Ichioka, and K. Machida, Eur. Phys. J. B \textbf{27}, 303
(2002); J. Phys. Soc. Jpn. \textbf{73}, 2049 (2004).
%
\bibitem{PJHirschfeld:1996} P. J. Hirschfeld and W. O. Putikka,
Phys. Rev. Lett. {\bf 77}, 3909 (1996).
%
\bibitem{takeya} J. Takeya {\sl et al.}, Phys. Rev. Lett. {\bf 88},
077001 (2002).
\end{thebibliography}
\end{document}